\documentclass[preprint,showpacs,preprintnumbers,amsmath,amssymb]{revtex4}

\def\be{\begin{eqnarray}}
\def\ee{\end{eqnarray}}
\def\ba{\begin{array}}
\def\ea{\end{array}}

\begin{document}



\title{The Interrelation between Incompressible Strips and \\
  Quantized Hall Plateaus} 

\author{AFIF SIDDIKI}
\author{ROLF R. GERHARDTS}
\affiliation{Max-Planck-Institut f\"ur Festk\"orperforschung, \\
Heisenbergstrasse 1, D-70569, Federal Republic of Germany \\
a.siddiki@fkf.mpg.de}


\received{23 June 2004}

\begin{abstract}
We study the current and charge distribution of a two dimensional electron
gas under strong perpendicular magnetic fields within the linear response 
regime.
We show within a self-consistent screening theory that incompressible 
strips with integer values of local Landau-level filling factor 
exist for finite intervals of the magnetic field strength $B$. Within an 
essentially local conductivity model, we find that the current density in 
these $B$ intervals is confined to the incompressible strips of vanishing 
local longitudinal resistivity. This leads to vanishing longitudinal and 
exactly quantized Hall resistance, and to a nice agreement of the 
calculated Hall potential profiles with the measured ones. 
\end{abstract}


\maketitle

\section{Introduction}
\vspace*{0.7cm}
Recent investigations\cite{Ahlswede01:562} with a scanning force microscope
 produced evidence,
 that the spatial distribution of the dissipative
current, carried by a two-dimensional electron gas (2D EG) in a Hall bar under
the conditions of the quantized Hall effect (QHE), is dominated by the
existence of incompressible strips (ISs), which are expected to develop in
the space regions where an integer number of Landau levels (LLs) is 
filled.\cite{Chklovskii92:4026} Subsequent theoretical
 work,\cite{Guven03:115327}  which calculated electron density and
electrostatic potential within a self-consistent Thomas-Fermi-Poisson
approximation, and current density and position-dependent
electrochemical potential from a local version of Ohm's law, confirmed
the experimental evidence and found that the current flows in the
ISs, in which the local longitudinal resistivity vanishes.
Due to the Thomas-Fermi approximation (TFA), incompressible strips were
found for all values $B$  of the magnetic field below a threshold
$B_c$, at which the center of the sample becomes
incompressible.\cite{Guven03:115327}
This leads to a  vanishing total resistance for $B<B_c$, to a
 potential drop in several steps across ISs, which may coexist near  a
 sample edge, and, for $B<B_c$, to the absence of $B$ intervals with
 linear variation of the Hall potential across the sample. These
 shortcomings are not in agreement with experiment\cite{Ahlswede01:562}
and require an improvement of the model.\cite{Guven03:115327}

\section{Improvements of the G\"uven-Gerhardts model}
We have recently improved on the approach of G\"uven and 
Gerhardts\cite{Guven03:115327}
in two essential respects\cite{Siddiki:preprint}. First, replacing the TFA 
by a Hartree-type
approximation, which takes the finite extent of wavefunctions into
account, we demonstrate that the existence of incompressible strips is
restricted to finite $B$ intervals below $B_c$\cite{Siddiki:preprint}. 
Second, we relax the assumption of a strictly local relation between
conductivity and electron density, which leads to singular current
density even along isolated lines of integer LL-filling factors, and
we simulate non-local effects  by a suitable averaging of the local
conductivity tensor\cite{Siddiki:preprint}.

Following the previous work,\cite{Oh97:13519,Guven03:115327} we
calculate density profile and effective confinement potential of the
2D EG in thermal equilibrium self-consistently. Assuming that all
charges, including a 
homogeneous positive background and induced charges on gates, defining
the  Hall bar (in $|x|<d$), are concentrated on the plane $z=0$ with
translation 
invariance in $y$-direction, we calculate the potential $V(x)$ from
the electron density $n_{\rm el}(x)$ solving Poisson's equation under
suitable boundary
conditions.\cite{Chklovskii92:4026,Oh97:13519,Guven03:115327}   
Electrostatic self-consistency requires that the density of the
spin-degenerate 2D EG is calculated from
\be n_{\rm el}(x)=\frac{2}{2\pi l^2} \int dX_0
\sum_{n=0}^{\infty} \frac{|\phi_{n,X_{0}}(x)|^{2}} 
{\exp  \big([E_{n}(X_{0})-\mu^{\star}]/k_BT\big)+1} \,,\label{geneled} \ee
where $l=\sqrt{\hbar/m\omega_{c}}$ is the magnetic length, $X_0$ a
center coordinate, 
$\mu^{\star}$ the electrochemical potential. 
The energy eigenvalues $E_{n}(X_{0})$ and eigenfunctions
$\phi_{n,X_{0}}(x)$ should be calculated from the single-electron
Schr\"odinger equation with the confinement potential $V(x)$.
In the previous work one assumed $V(x)$ to vary slowly on the scale $l$
and evaluated  Eq.(\ref{geneled}) in the TFA,
$E_{n}(X_{0})\approx E_{n}+V(X_{0})$, $|\phi_{n,X_{0}}(x)|^{2}\approx
\delta(x-X_{0})$, neglecting the extent of the
wavefunctions.\cite{Chklovskii92:4026,Oh97:13519,Guven03:115327} 
\begin{figure}[ht]
$ \left. \right. $ \includegraphics{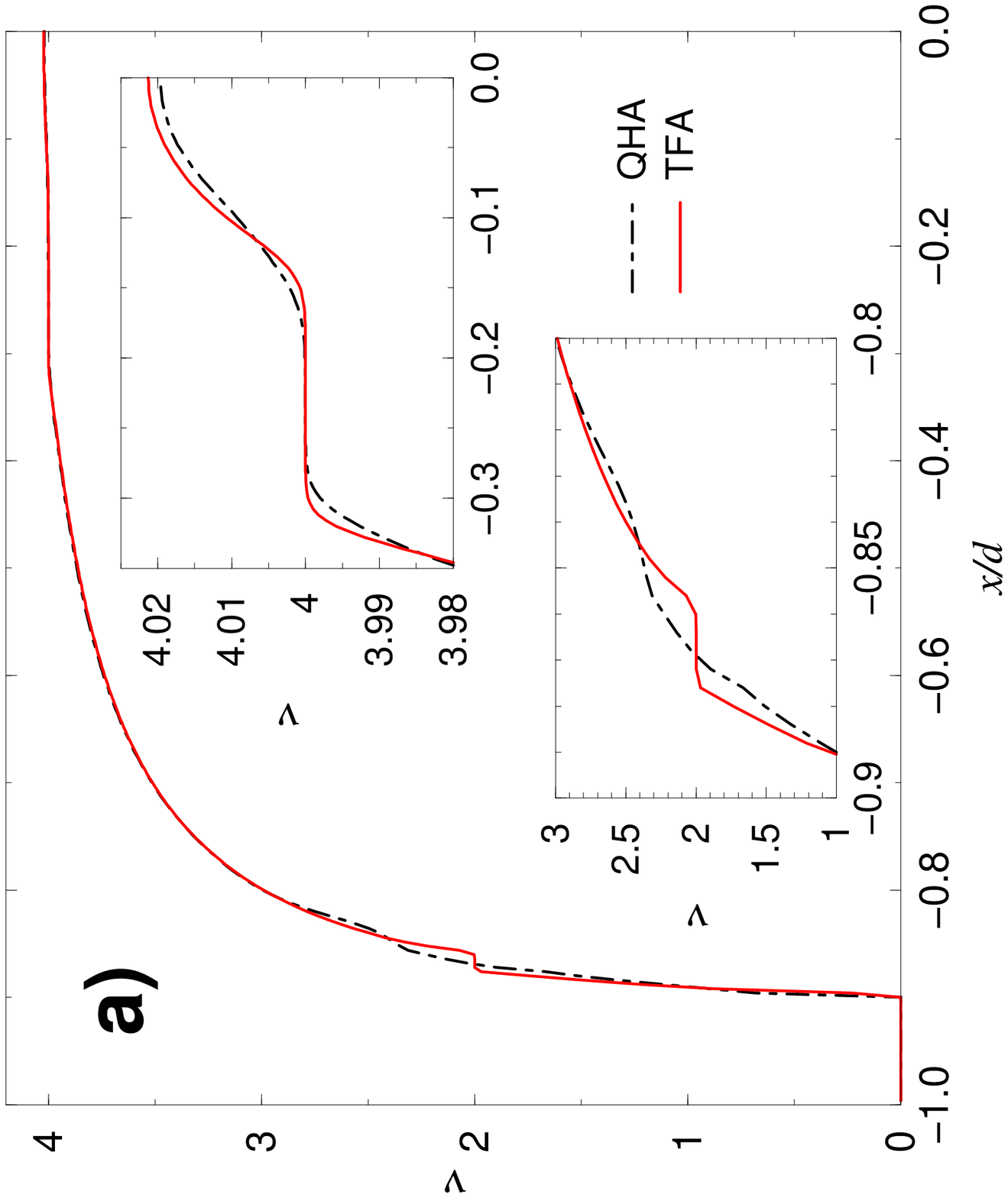}
$ \left. \right. $ \includegraphics{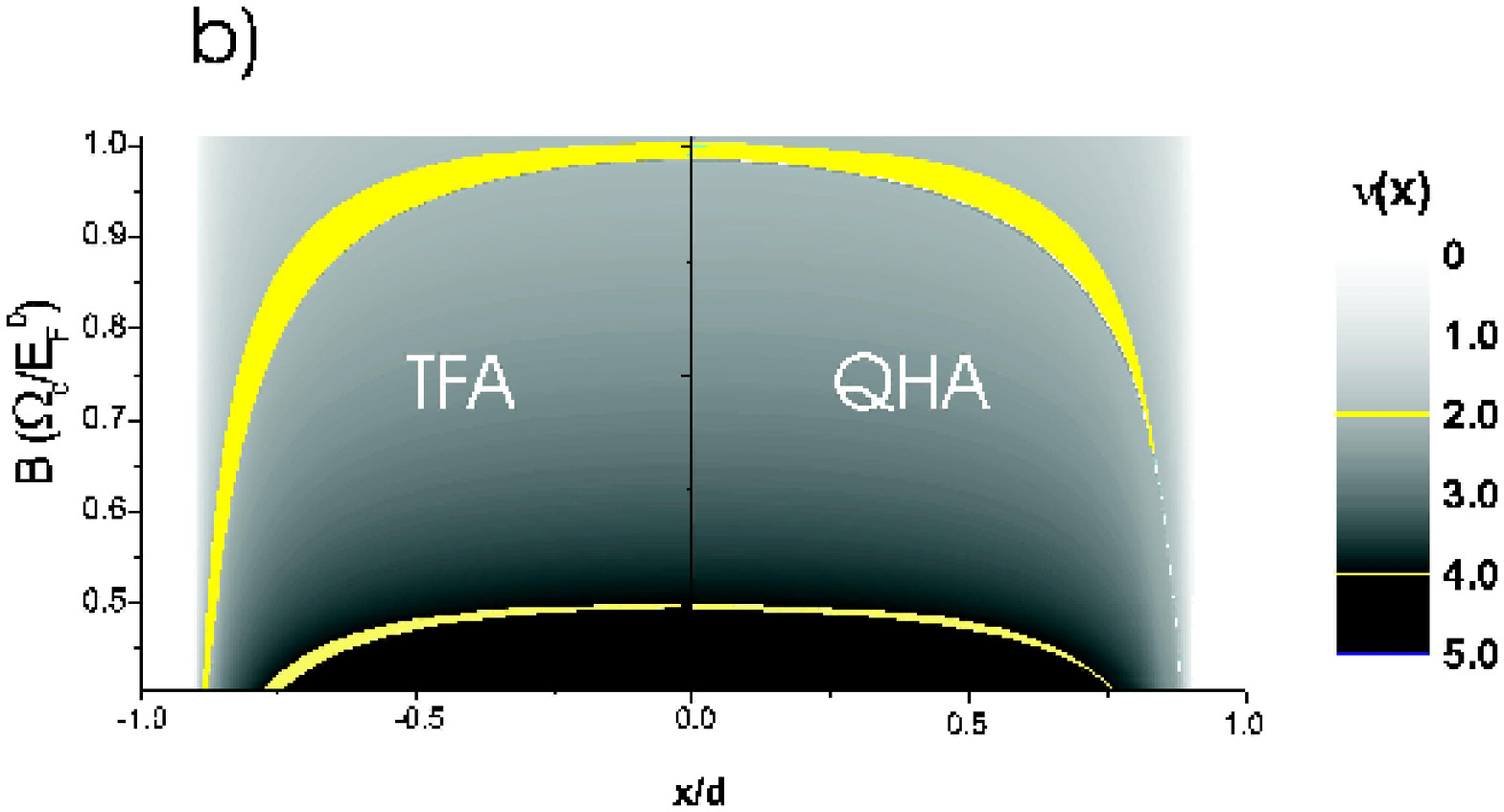} \vspace{6.5cm}
\caption{\label{fig1}(color online) (a) Calculated density profiles
  for the left half of a symmetric Hall bar. While the TFA yields
  incompressible strips with $\nu(x)=2$ and $\nu(x)=4$, only those
  with  $\nu(x)=4$ survive in the QHA.
 (b) Local filling factors (gray scale plot) versus
  position $x$ and cyclotron energy $\Omega_c \equiv \hbar
  \omega_c$. ISs (light) exist in the TFA for all $B$ 
  with $\Omega_{c}/E_{F}^{0}<1$, in the QHA, however, only for 
$1.0>\Omega_{c}/E_{F}^{0}>0.65$ and $0.5>\Omega_{c}/E_{F}^{0}>0.4$. 
 $d=3.1 \mu$m, average electron density  
$3\cdot10^{11}$cm$^{-2}$, temperature $k_{B}T/E_{F}^{0}=0.02$,
  $E_{F}^{0}=n_{\rm el}^0(0)\pi\hbar^{2}/m$, where $n_{\rm
  el}^0(0)= n_{\rm el}(0)$ 
at $T=0$, $B=0$.  } \end{figure}
This yields flat ISs in the density profile with
local filling factor $\nu(x)\equiv 2\pi l^2\,n_{\rm el}(x) =2$ for all
magnetic-field values $B<B_2\, (\equiv B_c)$, with $\nu(x)=4$ for all $B<B_4\,(
=B_2/2)$, etc. This is, however, an artifact of the TFA. If one takes
the width $\lambda_W$
of the wavefunctions into account, e.g. in the full Hartree
approximation, one finds that the ISs calculated in
the TFA  are smeared out if their width becomes smaller than
$\lambda_W$. This is 
already seen for the ``quasi-Hartree approximation'' (QHA), which uses
again $E_{n}(X_{0})\approx E_{n}+V(X_{0})$ but replaces the
$\phi_{n,X_{0}}(x)$ by the Landau wavefunctions of the unconfined 2D
EG, as is demonstrated in Fig.\ref{fig1}a and in Fig.\ref{fig1}b, which
shows within the QHA no ISs for $B_4\equiv 0.5B_2 <B < 0.65B_2$.

The stationary state with an applied dissipative current
$I$ was described\cite{Guven03:115327} by a local Ohm's law
$ {\bf E(r)}=\hat{\rho}({ \bf r}){\bf j(r)}$, with $\hat{\rho}({ \bf
r})=[\hat{\sigma}(n_{\rm el}{\bf r})]^{-1}$, which allows us to
calculate current density ${\bf j(r)}$ and driving  field ${\bf
  E(r)}=\nabla \mu^{\star}({\bf r})/e$ with the constant components $j_{x} 
\equiv 0$ and  $E_{y}(x)\equiv E_{y}^{0}$, and with
$j_{y}(x)=E_{y}^{0}/\rho_{l}(x)$, and 
$E_{x}(x)=E_{y}^{0}\rho_{H}(x)/\rho_{l}(x)$, with $\rho_{l}(x)$ the
longitudinal and $\rho_{H}(x)$ the Hall resistivity. This yields the 
Hall resistance  $R_{H}=V_{H}/I$ and the longitudinal resistance
$R_{l}=2dE_{y}^{0}/I$:
\be \label{equ2}
I=E_{y}^{0}\, \int_{-d}^{d}dx\, 1/\rho_{l}(x)\,, \quad 
 V_{H}= E_{y}^{0}\int_{-d}^{d} dx\,\rho_{H}(x)/\rho_{l}(x)\,. \ee
The feedback of the applied current on the (measurable) change of the
electrostatic potential\cite{Ahlswede01:562} was considered by
requiring the stationary state to be in local equilibrium.\cite{Guven03:115327}
Here we consider only a linear response situation, in which the
current-induced change of the electrostatic potential equals that of
the electrochemical potential. To avoid spurious singularities of the
integrals in Eq.(\ref{equ2}) along isolated lines of vanishing
$\rho_{l}(x)$, i.e. integer $\nu(x)$, which are an artifact of the
strictly local model, we average the conductivity tensor, 
\begin{figure}[ht] 
$ \left. \right. $ \includegraphics{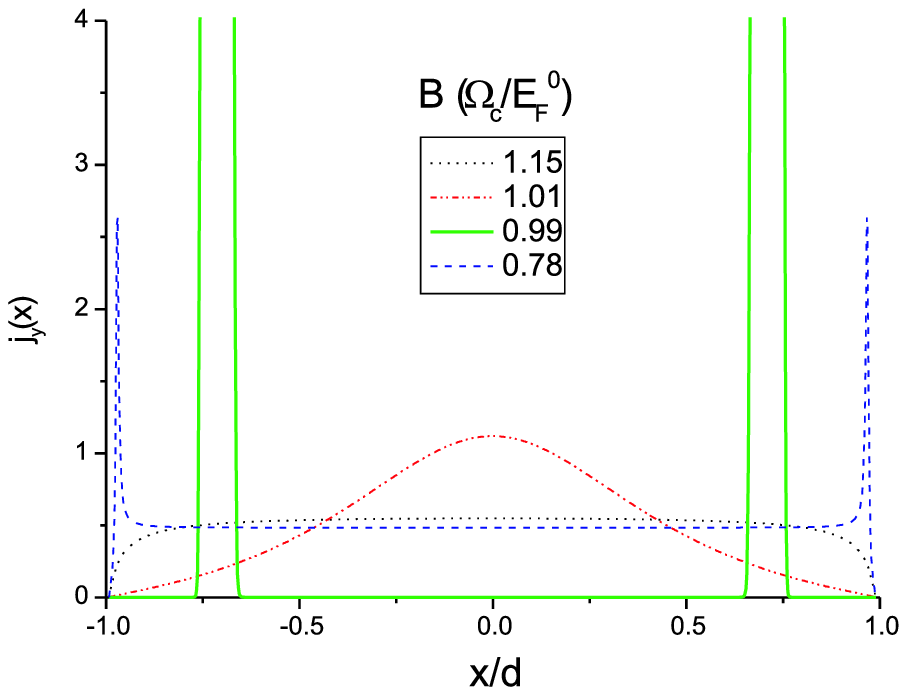}
$ \left. \right. $ \includegraphics{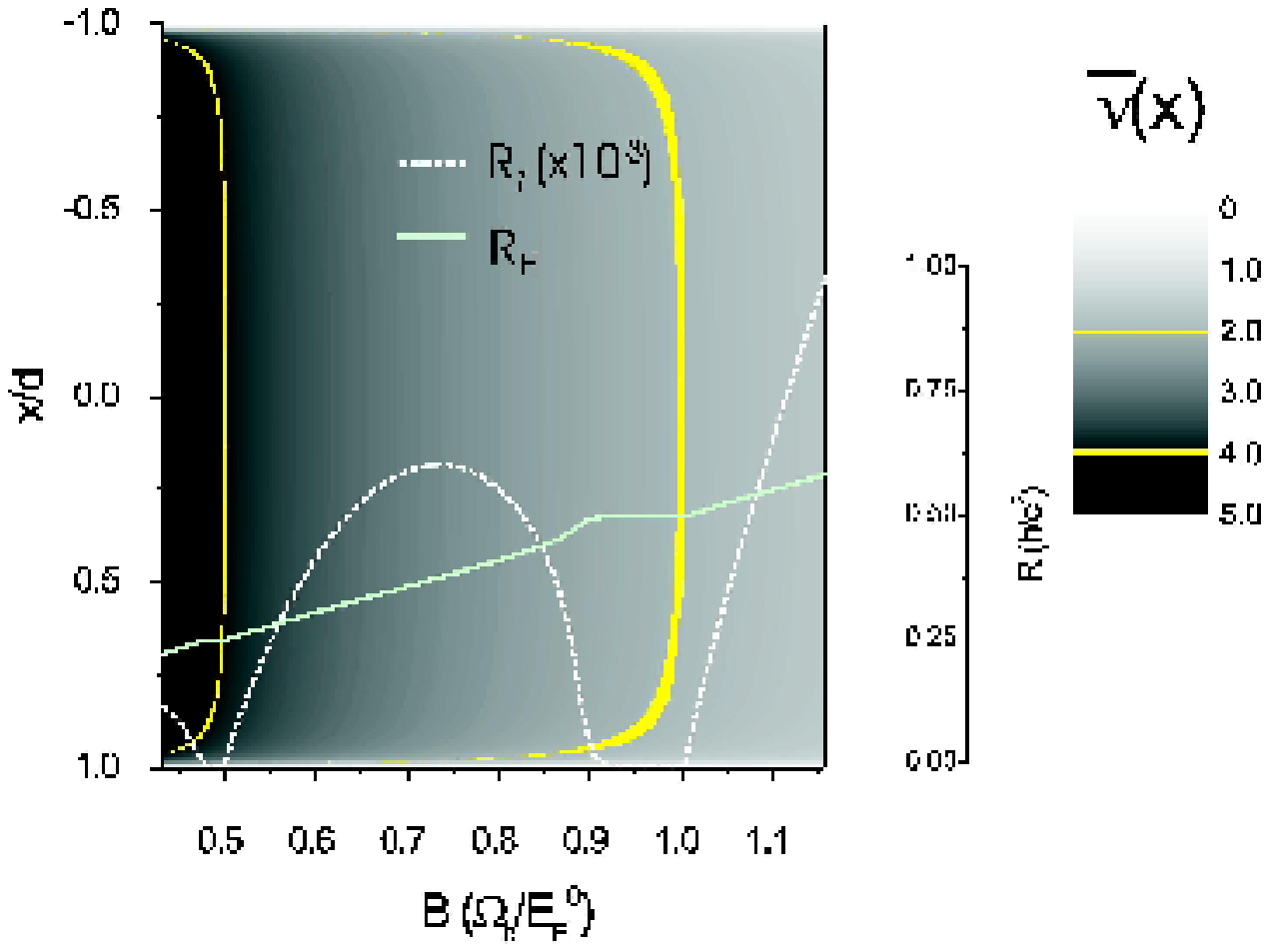} \vspace{6.5cm}
\caption{\label{fig2} (color online) (a) Current distribution for selected $B$
  values. Only for $\Omega_c/E_F^0=0.99$ (solid line) the current is
  confined to the incompressible strips, which survived the
  averaging. $\lambda=20$nm. 
 (b) Gray scale plot of the averaged $\bar{\nu}(x)\approx (h/e^2)\,
  \bar{\sigma}_H(x)$   versus $B$ and $x$, together with Hall (solid)
  and longitudinal (broken light line) resistance; $d=7.5 
  \mu$m, $\lambda=20$nm, $k_{B}T/E_{F}^{0}=0.02$.  \label{fig:numean}}
\end{figure}
over a length $\lambda$ of the order of the mean electron distance
(Fermi wavelength):
$\, \hat{\bar{\sigma}}(x)=\frac{1}{2\lambda}\int_{-\lambda}^{\lambda}
d\xi\, \hat{\sigma}(x+\xi).$

 Starting with the self-consistent Born approximation\cite{Ando82:437}
for the conductivity tensor of a homogeneous 2D EG,
$\hat{\sigma}(\nu)$, replacing  the filling factor $\nu$ by the local
 $\nu(x)$ calculated within the TFA, and performing the spatial
average, we find strips with 
vanishing denominators in Eq.(\ref{equ2}) only for ISs which are wider
than $2 \lambda$ and survive also in the QHA. If such wide ISs with
$\nu(x)=2k$ exist
in a certain $B$-interval, they carry all the current (cf. Fig.\ref{fig2}a)
and lead, for $T \rightarrow 0$,
 to exactly quantized resistance values,
$R_H=h/(e^2 2k)$, $R_l=0$. For all other $B$-values the current is
spread over the sample, and Eq.(\ref{equ2}) yields reasonable average
values for $R_H$ and $R_l$, as is demonstrated in Fig.\ref{fig2}b.
\section{Summary}
\begin{figure}[ht]
$ \left. \right. $ \includegraphics{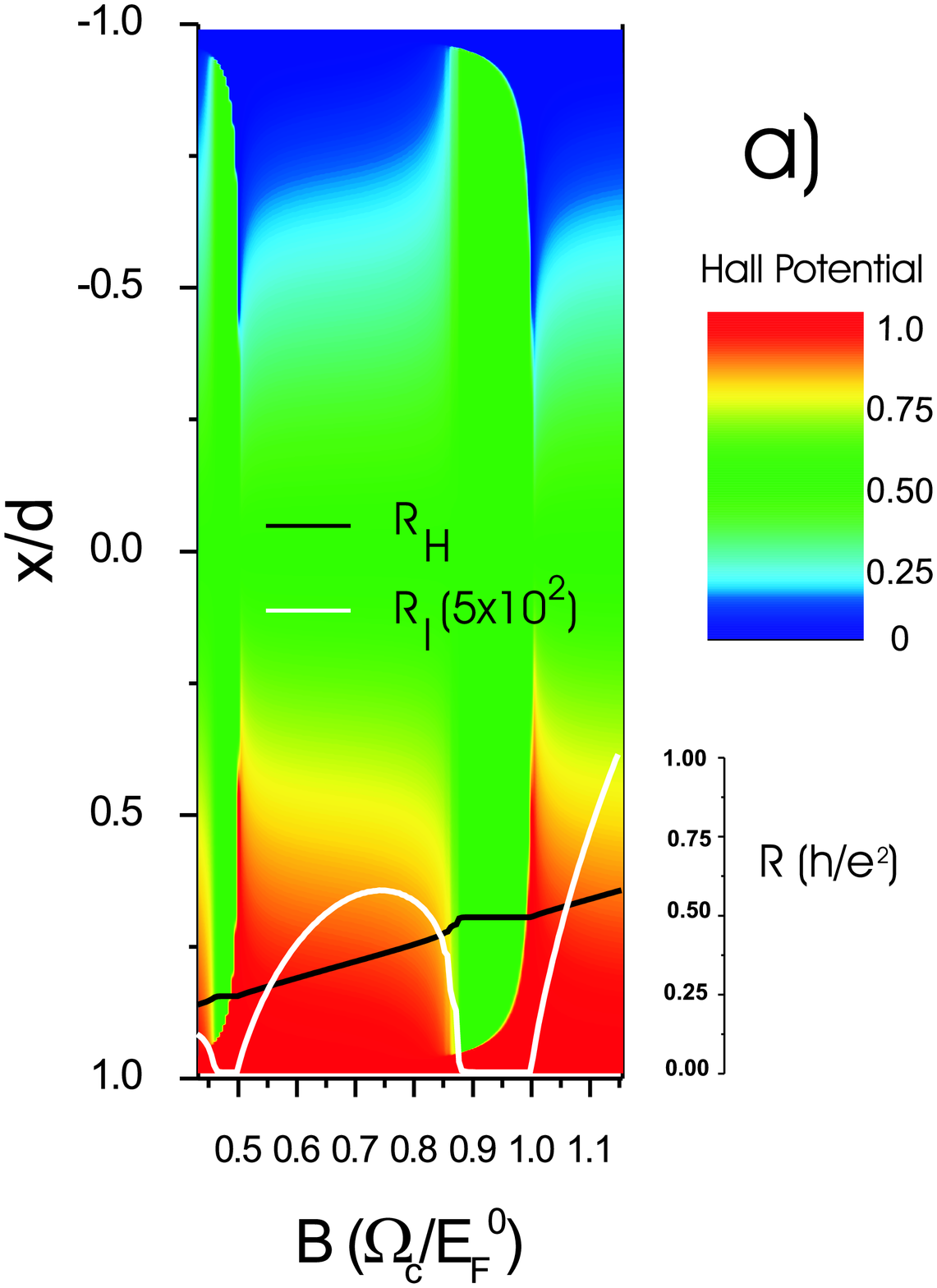}
$ \left. \right. $ \includegraphics{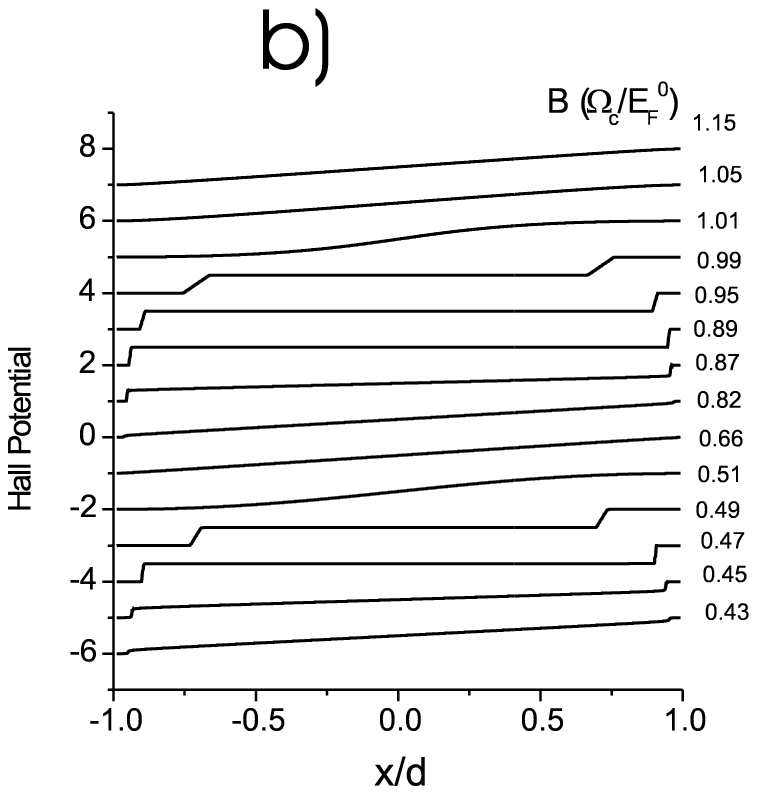} \vspace{7cm}
\caption{(color 
online) (a) Plot of the scaled Hall potential
  $\Phi(x)/\Phi(d)$, $\Phi(x)=\int_{-d}^x d\xi\, \bar{\sigma}_H(\xi) /
  \bar{\sigma}_l(\xi)$ versus $B$, together with $R_H(B)$ (black) and
  $R_l(B)$ (white line). $d=7.5\mu$m,
  $kT/E_{F}^{0}=0.0124$. \label{fig3}
 (b)  $\Phi(x)/\Phi(d)$ for selected values of
  $B$, vertically offset [cuts through (a)]. A stepwise
  variation is seen for $B$ in a plateau of the QHE, a
  linear variation if $B$ is far from a plateau, and a nonlinear
  variation in the center for $B$ near a plateau edge.
}
\end{figure}

In conclusion, the spatial averaging of the conductivity tensor, 
which simulates non-local effects and eliminates spurious
singularities due to narrow ISs, which are an artifact of the TFA and
do not survive in a Hartree-type approximation, represents  an 
essential improvement of the
theory and allows us to calculate reasonable results for the
longitudinal ($R_l$) 
and the Hall ($R_H$) resistance as continuous functions of $B$, with 
 high-precision plateau values   in
the $B$ intervals in which sufficiently wide ISs exist. Furthermore, we
find three different types of the variation of the Hall potential
across the sample (Fig.\ref{fig3}), in excellent agreement with the 
experiments.\cite{Ahlswede01:562}
\vspace*{-0.3cm}
\acknowledgments
We acknowledge useful discussions with E.~Ahlswede and J.~Weis, and
support by the Deutsche Forschungsgemeinschaft, SP
``Quanten-Hall-Systeme'', GE306/4-2.


\vspace*{-0.5cm}



\begin{thebibliography}{10}

\bibitem{Ahlswede01:562}
E. Ahlswede, P. Weitz, J. Weis, K. v.~Klitzing,  K. Eberl, Physica B {\bf
  298},  562  (2001).

\bibitem{Chklovskii92:4026}
D.~B. Chklovskii, B.~I. Shklovskii, and L.~I. Glazman, Phys. Rev. B {\bf 46},
  4026  (1992).

\bibitem{Guven03:115327}
K. G{\"u}ven and R.~R. Gerhardts, Phys. Rev. B {\bf 67},  115327  (2003).

\bibitem{Siddiki:preprint}
A. Siddiki and R.~R. Gerhardts, cond-mat/0406199 (unpublished).

\bibitem{Oh97:13519}
J.~H. Oh and R.~R. Gerhardts, Phys. Rev. B {\bf 56},  13519  (1997).

\bibitem{Ando82:437}
T. Ando, A.~B. Fowler, and F. Stern, Rev. Mod. Phys. {\bf 54},  437  (1982).

\end{thebibliography}
\end{document}